\documentclass[12pt]{iopart}
\expandafter\let\csname equation*\endcsname\relax
\expandafter\let\csname endequation*\endcsname\relax
\usepackage{amsfonts} 
\usepackage{amsmath}
\usepackage{iopams}
\usepackage{amssymb}
\usepackage{mathtools}
\usepackage{cancel} 
\usepackage{bm}
\usepackage[usenames, dvipsnames]{color} 
\usepackage{dcolumn}
\usepackage{epic} 
\usepackage{epsfig}
\usepackage{feynmp}
\usepackage{graphicx}
\usepackage{grffile}
\usepackage[breaklinks,colorlinks = true,linkcolor = red,urlcolor=blue,citecolor=red]{hyperref}
\usepackage{mathrsfs}
\usepackage{soul}
\usepackage{wrapfig}
\usepackage{xy} 
\usepackage{xcolor}
\usepackage{amsmath,amssymb,amsfonts,amsthm}
\usepackage[ascii]{inputenc}
\usepackage{makecell}
\allowdisplaybreaks

\newcommand{\be}{\begin{equation}}
	\newcommand{\ee}{\end{equation}}
\newcommand{\ba}{\begin{aligned}}
	\newcommand{\ea}{\end{aligned}}
\newcommand{\bea}{\begin{eqnarray}}
	\newcommand{\eea}{\end{eqnarray}}	
	
\newcommand{\beal}{\begin{align}}
\newcommand{\eal}{\end{align}}

\graphicspath{{images/}{../Figures}}

\begin{document}

\title[An exact formula for the variance of linear statistics in the $1d$- jellium model]
{An exact formula for the variance of linear statistics in the one-dimensional jellium model}

\author{Ana Flack}
\address{LPTMS, CNRS, Univ.  Paris-Sud,  Universit\'e Paris-Saclay,  91405 Orsay,  France}
\author{Satya N. Majumdar}
\address{LPTMS, CNRS, Univ.  Paris-Sud,  Universit\'e Paris-Saclay,  91405 Orsay,  France}
\author{Gr\'egory Schehr}
\address{Sorbonne Universit\'e, Laboratoire de Physique Th\'eorique et Hautes Energies, CNRS UMR 7589, 4 Place Jussieu, 75252 Paris Cedex 05, France}



\begin{abstract}
We consider the jellium model of $N$ particles on a line confined in an external harmonic potential and with a pairwise one-dimensional
Coulomb repulsion of strength $\alpha > 0$. Using a Coulomb gas method, we study 
the statistics of $s = (1/N) \sum_{i=1}^N f(x_i)$ where $f(x)$, in principle, is an arbitrary smooth function. While the mean of $s$ is easy to compute, the variance is nontrivial due to the long-range Coulomb interactions. In this paper we demonstrate that the fluctuations around this mean are Gaussian with a variance ${\rm Var}(s) \approx b/N^3$
for large $N$. We provide an exact compact formula for the constant $b = 1/(4\alpha) \int_{-2 \alpha}^{2\alpha} [f'(x)]^2\, dx$. In addition, we also calculate the full large deviation function characterising the tails of the full distribution ${\cal P}(s,N)$ for several different examples of $f(x)$. Our analytical predictions are confirmed by numerical simulations.  
\end{abstract}
\date{\today}
\maketitle

\maketitle
\section{Introduction}\label{Introduction}

Understanding the probability distribution of the sum of random variables is a well known problem in probability theory with
multiple applications in physics, chemistry, statistics and biology. Consider for instance $N$ random variables $\{x_1, x_2, \cdots, x_N\}$
drawn from a joint probability distribution function (jPDF) $P_{\rm joint}(x_1, x_2, \cdots, x_N)$. The general question is: how is the
sum $S_N = \sum_{i=1}^N x_i$ distributed? In the simplest case where the variables are independent and identically distributed (IID), each
drawn from $p(x)$ (with zero mean for simplicity), then the jPDF factorizes, i.e., $P(x_1, x_2, \cdots, x_N) = p(x_1) p(x_2) \cdots p(x_N)$. In this case, the central limit theorem (CLT) guarantees that, as long as the variance $\sigma^2 = \int x^2\, p(x)\,dx$ is finite, the sum $S_N$ converges for large $N$ 
to a Gaussian random variable, i.e., $S_N/(\sigma \sqrt{N}) \to  {\cal N}(0,1)$ where ${\cal N}(0,1)$ is a standard normal variable with zero mean and unit variance.

However, for correlated random variables, in particular when the correlations are strong, the CLT does not hold in general and determining the PDF
of $S_N$ for large $N$ is a challenging problem and appears in many different contexts. A celebrated example is a long-ranged gas with repulsive pairwise interactions and confined in a harmonic potential in one-dimension. This is the so-called Riesz gas where the energy of a configuration of charges with positions $x_i$'s on a line is given by \cite{riesz,lewin}
\bea \label{E_riesz}
E[\{ x_i \}] = \frac{A}{2}\,\sum_{i=1}^N  x_i^2 + \alpha \,{\rm sgn}(k) \sum_{i \neq j} |x_i - x_j|^{-k}  \;,
\eea
where $\alpha>0$ and $A>0$ are fixed coupling constants of order $O(1)$. While the first term represents the potential energy due to the external harmonic potential, the second term represents the
repulsive interaction between the $i$-th and $j$-th particle. In Eq. (\ref{E_riesz}), we assume that $k > -2$ (otherwise the harmonic potential is not sufficient to confine the gas). The jPDF of the positions of the particles is given by the Gibbs-Boltzmann distribution 
\bea \label{gibbs}
P(x_1, x_2, \cdots, x_N) = \frac{1}{Z_N} \, e^{- E[\{ x_i\}]/(k_B T)} \;,
\eea
where $T$ is the temperature, $k_B$ is the Boltzmann constant and $Z_N$ is the normalising partition function. In this case, the jPDF $P(x_1, x_2, \cdots, x_N)$ does not factorise because of the interaction term in Eq. (\ref{E_riesz}), which is long-ranged. Hence, it represents an example of a strongly correlated gas which has been extensively studied \cite{leble2017,leble2018,agarwal2019harmonically,jit2021,jit2022,santra2022}.  
There are three limiting cases of this Riesz gas that have been studied widely in the literature, namely (i) $k = -1$: this is the famous jellium model also known as the one-dimensional one-component plasma (1$d$-OCP)  \cite{lenard61,prager62,baxter1993statistical,aizenman80,dhar2017exact,dhar2018extreme,Flack2021,Flack2022,Chafai22}, (ii) $k \to 0^+$: this is the well known Dyson's log-gas \cite{Dyson1962,Dyson1963,Mehtabook,Forrester,Saffbook,majumdar2014top} where the positions $x_i$'s correspond to the eigenvalues of Gaussian random matrices and (iii) $k=2$, corresponding to the Calogero-Moser model which represents an important integrable classical model~\cite{calogero1971solution,calogero1975exactly,sutherland,moser,agarwalCM}.  In these examples of the Riesz gas, $S_N/N$ represents the position of the center of mass of the particles, a natural and important physical observable.

More generally, one can also study the probability distribution of the so-called linear statistics
\bea \label{def_lin_stat}
S_N = \sum_{i=1}^N f(x_i) \;,
\eea
where $f(x)$ is an arbitrary function (not necessarily linear). In the case when $f(x) = x$, it reduces to the center of mass (multiplied by $N$) of the gas. However, the case of a general $f(x)$ is also of interest and has been widely studied in several special cases, in particular in the context of the 
random matrix theory~\cite{Politzer,Beenaker1993B,Beenaker1993,Basor,Chen,baker,Joh,Soshnikov,Pastur2,sommers,Lytova,Cunden2014,Grabsch2,Grabsch3,Grabsch5,Texier13}. Some examples are as follows
\begin{itemize}
\item[$\bullet$]{full counting statistics in the Dyson's log-gas \cite{Soshnikov,CL,FS,index1,index2,marino14,smith} or in the jellium model \cite{dhar2018extreme,Flack2022}, where one is interested in the number of particles $N_{L} = \sum_{i=1}^N {\mathbb{I}}_{[-L,L]}(x_i)$ in an interval $[-L,+L]$. Here, $f(x) = {\mathbb{I}}_{[-L,L]}(x)$ is an indicator function, i.e., ${\mathbb{I}}_{[-L,L]}(x) = 1$ if $-L \leq x \leq L$ and zero otherwise.}
\item[$\bullet$]{conductance (with $f(x) = x$) and shot noise ($f(x) = x(1-x)$) in chaotic transport through a cavity where $0\leq x_i \leq 1$'s represent the eigenvalues of an $N \times N$ Jacobi random matrix \cite{Beenaker1993B,sommers,Cunden2014,kanzieper1,bohigas1,Khor,kanzieper2,bohigas2,kedar,Grabsch1,Grabsch4}.}
\item[$\bullet$]{R\'enyi entropy in a random pure state of a bipartite system where $f(x) = x^q$, with $q >0$ represents the R\'enyi index \cite{parisi,Nadal1,Nadal2}.}
\end{itemize}
In the case of Dyson's log-gas, i.e., $k \to 0^+$ limit of the Riesz gas, the linear statistics $S_N$ in Eq. (\ref{def_lin_stat}) with general $f(x)$ have been studied extensively and several exact results are known. In order to study the linear statistics in the log-gas, it is convenient to first rescale $x_i \to x_i \sqrt{N}$'s such that the average density of the gas  is supported, for large $N$, over an interval of size of order $O(1)$. Then one defines $s =(1/N) \sum_i f(x_i)$ where $x_i$'s now represent the rescaled coordinates such that $s \sim O(1)$ in the large $N$ limit. It is known that the distribution of the typical value of 
$s$ for a general $f(x)$ is Gaussian around its mean $\bar{s}$ with a variance given, for large $N$, by an explicit formula~\cite{Beenaker1993B,Joh,Cunden2014}
\bea \label{variance_RMT}
{\rm Var}(s) \approx \frac{1}{\beta \pi^2 N^2} \int_0^\infty \, k\, |\hat f(k)|^2 \, dk \;, 
\eea
where $\hat f(k) = \int_{-\infty}^\infty e^{i k x} f(x)\, dx$ is the Fourier transform of $f(x)$ and the parameter $\beta = 1, 2, 4$ represents the three standard symmetry classes of Gaussian random matrices. Note that this formula assumes that the integral in (\ref{variance_RMT}) is convergent. For certain choices of $f(x)$ one has to augment this formula with some regularisation. For example for the center of mass where $f(x) = x$, its Fourier transform $\hat f(k)$ is not well defined and one cannot use the formula in Eq. (\ref{variance_RMT}) directly and use alternative formulas that are more complicated \cite{Beenaker1993B}. 

A natural question is whether one can obtain an explicit formula for general $f(x)$ for other values of $k$ in the Riesz model (\ref{E_riesz}). The purpose of this paper is to derive an explicit formula for $k=-1$, i.e., for the $1d$ jellium model, in the limit of large $N$. As in the case of the log-gas discussed above, it is useful to first rescale
$x_i \to L_N\, y_i $ where $y_i = O(1)$ and $L_N$ is an $N$-dependent length scale to be chosen as follows. Under this rescaling, the energy in Eq. (\ref{E_riesz}) for $k=-1$ reads 
\bea \label{E_yi}
E[\{y_i \}] = \frac{A}{2} L_N^2 \sum_{i=1}^N y_i^2 - \alpha \, L_N \, \sum_{i\neq j} |y_i - y_j| \;.
\eea
For the system to exhibit any interesting physical behavior, the two terms in Eq. (\ref{E_yi}) must scale in the same way for large $N$. The first term scales as $A \,L_N^2\,N$ since the sum $\sum_i y_i^2 \sim N$ as $y_i = O(1)$. The second term scales as $\alpha\, L_N\, N^2$ since there are $N(N-1)$ terms of order $O(1)$ each in the double sum. Equating these two, one gets, for large $N$, 
\bea \label{eq:LN}
A\, L_N \sim \alpha \, N  \;.
\eea 
Since both the coupling constants $A$ and $\alpha$ are of order $O(1)$, this tells us that we must choose $L_N = O(N)$. Choosing $L_N = N$ and setting $A =1$ for convenience the energy in Eq. (\ref{E_yi}) can be written  as
\bea \label{E_OCP_yi}
 E[\{y_i\}] = \frac{N^2}{2}\sum_{i=1}^{N}y_i^2 - \alpha\, N\, \sum_{i\neq j}|y_i-y_j| \;.
\eea  
For convenience, we henceforth will denote $y_i \to x_i$ and write the energy as
\bea \label{E_OCP}
 E[\{x_i\}] = \frac{N^2}{2}\sum_{i=1}^{N}x_i^2 - \alpha\, N\, \sum_{i\neq j}|x_i-x_j| \;.
\eea  
Note that, since $x_i$'s are typically of order $O(1)$, both terms in the energy in (\ref{E_OCP}) scale as $N^3$. Hence the total energy scales as $N^3$. 
In this case, the average density is defined by
\bea \label{def_rho}
\langle \rho_N(x) \rangle = \frac{1}{N}  \Big\langle\sum_{i=1}^N \delta(x-x_i) \Big\rangle \;,
\eea
where $\langle \cdots \rangle$ denotes an average over the Gibbs-Boltzmann measure in Eq. (\ref{gibbs}). It 
is known to converge in the large $N$ limit to a flat profile supported over $[-2 \alpha, 2 \alpha]$~\cite{prager62, dhar2017exact}.
\bea \label{flat}
\lim_{N \to \infty} \langle \rho_N(x) \rangle = \frac{1}{4\alpha}  \quad, \quad -2\alpha \leq x \leq 2 \alpha \;.
\eea 
As in the log-gas case, we now define the linear statistics in the rescaled coordinates as
\bea \label{def_s}
s = \frac{1}{N} \sum_{i=1}^N f(x_i) \;,
\eea
such that $s \sim O(1)$ in the large $N$ limit. In this case, using the average density in (\ref{flat}), it is easy to see that 
the average value of $s$ converges to 
\bea \label{av_s}
\bar{s} = \frac{1}{4\alpha}\int_{-2\alpha}^{2\alpha} f(x)\,dx \;.
\eea
Our main result in this paper is to obtain, using a Coulomb gas and large deviation method, an explicit formula for the variance ${{\rm Var} (s)} = \langle (s-\bar{s})^2\rangle$ which simply reads for large $N$
\bea \label{var_s}
{{\rm Var} (s)} \approx \frac{1}{4 \alpha N^3} \int_{-2 \alpha}^{2 \alpha}  \left[f'(x)\right]^2 \, dx \;.
\eea
This formula is valid only when $[f'(x)]^2$ exists and is integrable.  
We will discuss precisely the assumptions implicit in deriving this explicit formula and also verify this prediction by numerical simulations in several specific examples of $f(x)$. Using the same Coulomb gas method, we show how to access the full large deviation function that characterizes the tails of the full distribution ${\cal P}(s,N)$. For some examples of $f(x)$, we compute explicitly the rate function describing these large atypical fluctuations. In some cases, such as $f(x) = |x|$, we show from explicit computation that the rate function has a singular point where the third derivative is discontinuous, signalling a third order phase transition in the underlying Coulomb gas.

We will also provide an alternative derivation of this formula (\ref{var_s}), which in addition demonstrates that, up to an overall constant, the formula in Eq. (\ref{var_s}) holds for a general external potential $V(x_i)$, not necessarily harmonic. 

Let us remark that the variance of linear statistics has also been studied in another well known ensemble of RMT, namely the Ginibre ensemble where the entries are complex Gaussian but there is no Hermitian symmetry. In this case, the eigenvalues are complex and the eigenvalues behave like charged particles in two-dimensions repelling each other via a logarithmic Coulomb interaction and also confined in a harmonic potential. For this $2d$-case, a formula for the variance of $s = (1/N) \sum^N_{i=1}f({\vec{r_i}})$ in the large $N$ limit was derived by Forrester \cite{forrester_ginibre}, which looks formally very similar to our one-dimensional formula in (\ref{var_s}), up to overall constants and $N$-dependent factors, i.e., ${\rm Var}(s) \propto \int (\nabla f)^2 d \vec{r}$ where the integral is over the support of the Coulomb gas. This result in $d=2$ was
later proved rigorously in the mathematics literature \cite{virag,ameur} and the formula was also extended to confined Coulomb gases in $d \geq 2$ \cite{serfaty,armstrong}. However, these rigorous methods do not seem to easily extend to the jellium model in $d=1$ for which our results show that 
this formula for the variance, up to an overall factor, holds even in $d=1$. Our method, in addition, gives access to the full large deviation function for any linear statistics.

The rest of the paper is organised as follows. In Section \ref{sec:II}, we discuss the Coulomb gas method leading to the exact asymptotic formula in Eq. (\ref{var_s}) for general $f(x)$. We also study few examples where we provide explicit results for the large deviation function associated with the full distribution of $s$. In Section \ref{sec:alter}, we provide an alternative derivation of the formula for the variance. In Section \ref{sec:validity}, we discuss the criteria for the validity of this general formula. Finally, in Section \ref{sec:final}, we conclude with a summary and open problems.

\section{The computation of the variance via a Coulomb gas method}\label{sec:II}

We consider the linear statistics $s$ defined in Eq. (\ref{def_s}) with an arbitrary function $f(x)$. Clearly $s$ is a random variable since
the $x_i$'s are also random variables distributed via the Gibbs-Boltzmann weight (\ref{gibbs}) with the energy given in Eq. (\ref{E_OCP}). 
Our goal is to calculate the variance ${\rm Var}(s)$ of $s$ in the large $N$ limit. To compute it, we use the following strategy. We first express
the PDF ${\cal P}(s,N)$ as follows
\be \label{eq:LS0}
\mathcal{P}(s, N)  =  \int dx_1 \int dx_2 \cdots \int dx_N {P}(\{x_i\})\delta\left(s-\frac{1}{N}\sum_{i=1}^{N}f(x_i)\right) \;,
\ee 
with ${P}(\{x_i\})$ defined in Eq.\eqref{gibbs} and \eqref{E_OCP}. The idea is to evaluate this distribution in the large $N$ limit using a
Coulomb gas method detailed below. For large $N$, we will see that $\mathcal{P}(s, N) $ admits a large deviation form 
\bea \label{def_psi}
\mathcal{P}(s, N)  \sim e^{-N^3 \Psi(s)} \;,
\eea
where $\Psi(s)$ is a rate function that implicitly depends on $f(x)$. The factor $N^3$ in the exponent arises from the fact that the energy scales as $N^3$. 
Typically, one would expect that $\Psi(s)$ has a minimum around $s = \bar{s}$ (where $\bar{s}$ denotes the average value of $s$ in Eq.~(\ref{av_s})) and behaves quadratically around its mean
\bea \label{psi_quad}
\Psi(s) \approx  \frac{1}{2b}(s-\bar{s})^2 \;.
\eea
Substituting this behaviour in the large deviation form (\ref{def_psi}) one gets
\bea \label{gauss}
\mathcal{P}(s, N)  \sim e^{-\frac{N^3}{2b}(s-\bar{s})^2} \;,
\eea
indicating that the distribution, near its peak, has a Gaussian form with mean $\bar{s}$ and variance 
\bea \label{rel_b_var}
{\rm Var}(s) \approx \frac{b}{N^3} \;.
\eea
Therefore the idea would be to compute first the large deviation function $\Psi(s)$ and read off the number $b$ by expanding $\Psi(s)$ up to 
quadratic order around its minimum at $s = \bar{s}$. In the next subsection, we outline the Coulomb gas method to compute $\Psi(s)$.

\subsection{The Coulomb gas method: general set up}\label{sub:Coulomb}

We start from the multiple integral in (\ref{eq:LS0}). The first step is to get rid of the delta-function by replacing it  
with its integral representation $\delta(x) = N^3 \int_{\Gamma} \frac{d\mu}{2\pi} e^{- \mu N^3 x}\, dx$ where $\Gamma$ is a Bromwich contour going along the imaginary axis in the complex $\mu$-plane. We can then rewrite (\ref{eq:LS0}) 
as (up to an overall constant scale factor $N^3$)
\begin{equation}\label{eq:full0}
    \mathcal{P}(s, N) = \frac{\int_{\Gamma}d\mu \int dx_1 \int dx_2 \cdots dx_N e^{- E_{\mu}[\{x_i\}]}}{\int dx_1 \cdots dx_N\, e^{-E[\{x_i\}]}} = \frac{\int_\Gamma d\mu\, Z_N(\mu)}{Z_N} \;.
\end{equation}
In the denominator, the energy function 
$E[\{x_i\}]$ is given in Eq. (\ref{E_OCP}) -- note that for convenience we have set $k_B T = 1$, without any loss of generality. In the numerator in Eq. (\ref{eq:full0}), the modified energy function $E_{\mu}[\{x_i\}] $ reads
\begin{eqnarray}\label{eq:ener2}
E_{\mu}[\{x_i\}] &=& \frac{N^2}{2}\sum_{i=1}^{N}  x_i^2 - N\alpha \sum_{i\neq j}|x_i-x_j|  + N^3\mu \left[\frac{1}{N}\sum_{i=1}^Nf(x_i) - s\right]  \nonumber \\
  &=& \frac{N^2}{2}\sum_{i=1}^{N} \left( x_i^2 + \mu \, f(x_i)\right) - N\alpha \sum_{i\neq j}|x_i-x_j|  - \mu \, s \, N^3\;.
\end{eqnarray}
Up to a constant shift of energy $- \mu \,s \, N^3$, the energy in Eq. (\ref{eq:ener2}) can be interpreted as the energy of the same jellium gas, but in the presence of an effective potential
\bea \label{Veff}
V_{\text{eff}}(x)= \frac{1}{2}x^2 + \mu \, f(x) \;.
\eea
Thus the ``chemical potential'' $\mu$ can be interpreted as the amplitude of the perturbation of the original potential $x^2/2$. Consequently, 
$Z_N(\mu = 0) = Z_N$ in Eq. (\ref{eq:full0}). 

Evaluating the multiple integrals in the numerator and denominator of Eq. (\ref{eq:full0}) for all $N$ is hard in general. However, for large $N$, one
can make a ``continuum/hydrodynamic'' approximation. This proceeds in two steps. In the first step, we fix a macroscopic density profile $\rho(x)$ (normalised to unity) and sum over all microscopic configurations of $x_i$'s that correspond to this macroscopic density profile. In the second step, we integrate (functional integration) over all possible (normalised to unity) macroscopic density profiles. Following these two steps, one can write $Z_N(\mu)$ in Eq. (\ref{eq:full0}) as (up to an overall $N$-dependent constant factor) \cite{dean08}
\be \label{Zmu_1}
Z_N(\mu) \approx \int {\cal D}\rho(x) e^{-N^3 E_\mu[\rho(x)] - N \int dx \, \rho(x) \ln \rho(x)} \delta\left( \int dx \rho(x) - 1\right) \;,
\ee
where $E_\mu[\rho(x)] $ is given by
\be \label{E_mu_rho}
E_\mu [\rho(x)] = \int \left( \frac{x^2}{2} + \mu f(x)\right)\,\rho(x)\,dx - \alpha \int \int \rho(x) \rho(y) |x-y|\, dx \, dy  - \mu\,s  \;.
\ee
The second term $N \int dx \, \rho(x) \ln \rho(x)$ inside the exponent in the integrand in Eq.~(\ref{Zmu_1}) is an entropy term that comes from the first step of coarse-graining mentioned above, i.e., from the sum over all possible microscopic configurations corresponding to a given macroscopic density profile $\rho(x)$. Note that this entropy term scales as $O(N)$, while the energy $\sim N^3$ is much bigger. Hence in the large $N$ limit, we will henceforth neglect the entropy term. The delta-function in Eq. (\ref{Zmu_1}) represents the fact that only the normalised (to unity) macroscopic density profiles are allowed. In fact, 
it is again convenient to replace this delta-function by its integral representation $\delta(x) = N^3 \int_{\Gamma} \frac{d\mu_0}{2\pi} e^{- \mu_0 N^3 x}\, dx$. 
Finally the distribution ${\cal P}(s,N)$ can then be written as
\begin{equation}\label{eq:P1_t} 
\mathcal{P}(s, N) \approx \frac{\int \mathcal{D}\rho(x) \;\int d\mu \int d\mu_0  \;e^{-N^3S[\rho(x), \mu, \mu_0]}}{\int \mathcal{D}\rho(x) \int d\mu_0 \;  e^{-N^3 S[\rho(x), \mu=0,\mu_0]}}.
\end{equation}
where the effective action is given by
\begin{align}\label{eq:act2B}
  S [\rho (x), \mu, \mu_0 ] = \int \frac{x^2}{2}\rho(x)dx -\alpha\int \rho(x)\rho(y)|x-y|dxdy + \nonumber \\ 
  +\mu\left(\int f(x)\rho(x)dx-s\right)  +\mu_0\left(\int\rho(x)dx-1\right) \;.
\end{align}
The next step is to evaluate both the numerator and the denominator in Eq. (\ref{eq:P1_t}) by the saddle point method valid for large $N$. We consider them separately, starting with the denominator. 

\vspace*{0.3cm}
\noindent{\bf Denominator.} In this case, the saddle point equations read
\be \label{eq:spe}
\frac{\delta S[\rho (x), \mu=0,\mu_0]}{\delta \rho (x)} = 0 \quad {\rm and} \quad \frac{\partial S[\rho (x), \mu=0,\mu_0]}{\partial \mu_0} =0 \;.
\ee
The second equation actually gives the normalisation condition $\int \rho(x) \, dx = 1$, while the first equation gives the saddle point density \cite{prager62,dhar2017exact}
\be 
\rho_{0}^*(x) = \frac{1}{4\alpha}, \quad -2\alpha \leq x\leq 2\alpha \;,
\ee
which is flat over the finite support $[-2 \alpha, +2 \alpha]$. Inserting the saddle-point density into \eqref{eq:act2B} with $\mu=0$, one gets the leading large $N$ behavior of the unconstrained partition function of the jellium model \cite{prager62, dhar2017exact}
\begin{equation}\label{eq:partition_f1}
    Z_N = Z_N(\mu=0)\approx e^{\frac{2}{3}\alpha^2 \, N^3} \;.
\end{equation}

\vspace*{0.3cm}
\noindent{\bf Numerator.} In the case of the numerator, there are three saddle-point equations
\be \label{eq:spe_mu}
\frac{\delta S[\rho (x), \mu,\mu_0]}{\delta \rho (x)} = 0 \quad , \quad \frac{\partial S[\rho (x), \mu,\mu_0]}{\partial \mu} =0 \quad {\rm and} \quad 
\quad \frac{\partial S[\rho (x), \mu,\mu_0]}{\partial \mu_0} =0 \;.
\ee
The last two equations give the two constraints: $\int f(x) \rho(x) \, dx = s$ and $\int \rho(x)\, dx = 1$. We therefore need to find the solution $\rho_\mu^*(x)$ of the first equation that satisfies these two constraints. The first equation in (\ref{eq:spe_mu}), using the action $S[\rho (x), \mu,\mu_0]$ from (\ref{eq:act2B}), reads
\be \label{eq:derivative}
 \frac{x^2}{2}-2\alpha\int \rho^*_{\mu}(y)|x-y|dy + \mu f(x) + \mu_0 = 0 \;.
\ee
We assume that the saddle-point density $\rho_\mu^*(x)$ has a single support $x \in [L_1, L_2]$ (to be determined a posteriori). This equation holds only for $x$ belonging to the support. Taking a derivative of Eq. (\ref{eq:derivative}) with respect to $x$ gives
\be \label{eq:first_x}
x-2\alpha \int\rho^*_{\mu}(y){\rm sgn}(x-y)dy + \mu f'(x) = 0 \;.
\ee
Taking further one more derivative and using $\frac{d}{dx} {\rm sgn}(x) = 2 \delta(x)$ one finds
\be \label{eq:second_xB}
1+\mu f''(x) -4\,\alpha\rho^*_{\mu}(x)=0 \quad {\rm implying} \quad \rho_\mu^*(x) = \frac{1}{4 \alpha} \left( 1 + \mu f''(x)\right) \;, \; L_1 \leq x \leq L_2 \;.
\ee
\begin{figure}[t]
\centering
\includegraphics[width = 0.7\linewidth]{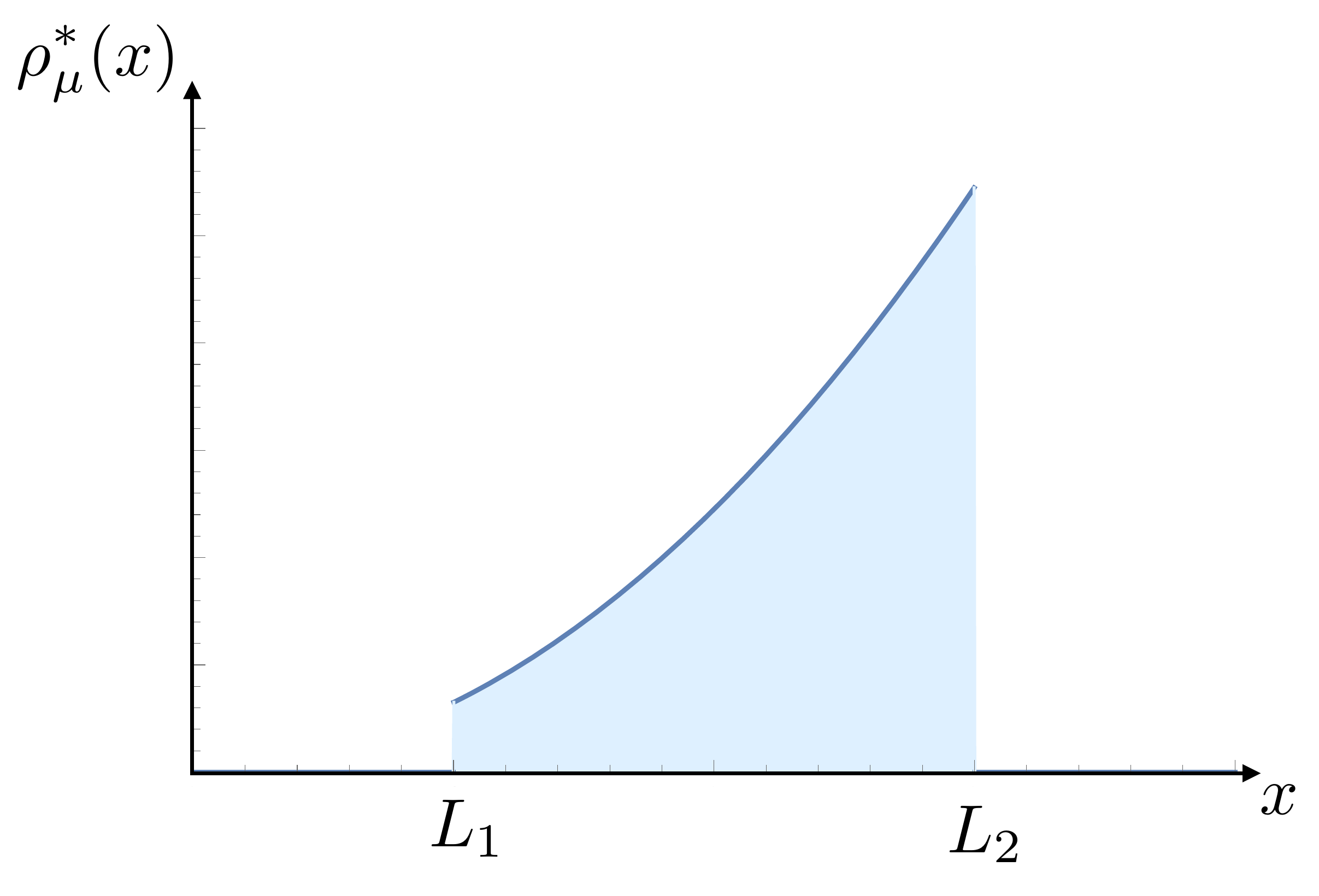}
\caption{A schematic plot of the density $\rho_\mu^*(x)$ in Eq. (\ref{eq:second_xB}) for $\alpha=1$, $\mu=1$ and $f(x) = x^4$. The left and the right edges of the support are denoted respectively by $L_1$ and $L_2$.}\label{Fig_rhomustar}
\end{figure}
For a schematic plot of this density in Eq. (\ref{eq:second_xB}) -- with $f(x) = x^4$ as an example -- see Fig. \ref{Fig_rhomustar}. 
At this stage, we have four unknown parameters: $L_1$, $L_2$, $\mu_0$ and $\mu$. They can determined as follows. Setting $x=L_2$ in Eq. (\ref{eq:first_x}) and using ${\rm sgn}(L_2 -y) = 1$ for all $y \leq L_2$ and using the normalisation $\int_{L_1}^{L_2} \rho_\mu^*(x) \, dx = 1$, one gets
\begin{equation}\label{eq:L2}
    L_2-2\alpha +\mu f'(L_2)=0 \;.
\end{equation}
Similarly, by setting $x=L_1$ in in Eq. (\ref{eq:first_x}) and using ${\rm sgn}(L_1 -y) = -1$ for all $y \geq L_1$ gives 
\begin{equation}\label{eq:L1}
    L_1+2\alpha +\mu f'(L_1) =0 \;.
\end{equation}
Setting $x=L_2$ in Eq. (\ref{eq:derivative}) and using $|L_2-y| = L_2-y$ for all $y \leq L_2$, gives a third independent relation
\be \label{eq:mu0_2}
\mu_0 = -\frac{L_2^2}{2}+2\alpha L_2-\frac{1}{4}(L_2^2-L_1^2)-\frac{\mu}{2}\int_{L_1}^{L_2}yf''(y)dy - \mu f(L_2) \;.
\ee  
Note that we have already used the normalisation condition in arriving at the first two equations (\ref{eq:L2}) and (\ref{eq:L1}). But we are still left with one condition
\bea \label{eq:condition1}
\int_{L_1}^{L_2} f(x) \rho^*_\mu(x) \, dx = s \;.
\eea
Inserting the saddle-point density from Eq. (\ref{eq:second_xB}) in this condition and performing the integral gives the desired fourth relation
\bea
    \label{eq:third}
    \frac{\mu}{4\alpha}\int_{L_1}^{L_2}f(x)f''(x)dx = s-\frac{1}{4\alpha}\int_{L_1}^{L_2}f(x)dx \;.
\eea
Thus the four unknown parameters ($L_1, L_2, \mu_0$ and $\mu$) are determined from the four independent non-linear relations (\ref{eq:L2}), (\ref{eq:L1}), (\ref{eq:mu0_2}) and (\ref{eq:third}). Once they are determined (for a given $s$), they characterise the saddle-point density $\rho_\mu^*(x)$ in Eq. (\ref{eq:second_xB}) fully. Let us remark that, when $\mu = 0$, one recovers the unconstrained density $\rho_0^*(x) = 1/(4 \alpha)$ for $L_1 \leq x \leq L_2$
with $L_1 = -2 \alpha$ and $L_2 = +2 \alpha$. In this case, we get $s = \bar{s}$ given in Eq. (\ref{av_s}).

Substituting the saddle-point density $\rho_\mu^*(x)$ back in Eqs. (\ref{eq:P1_t}) and (\ref{eq:act2B}) gives the leading large $N$ behavior of the numerator 
\bea \label{asympt_num}
\int \mathcal{D}\rho(x) \;\int d\mu \int d\mu_0  \;e^{-N^3\,S[\rho(x), \mu, \mu_0]} \sim e^{-N^3 S[\rho_\mu^*(x), \mu, \mu_0]} \;,
\eea 
where the saddle-point action $S[\rho_\mu^*(x), \mu, \mu_0]$ depends implicitly on $s$. Using the behavior of the denominator in Eq. (\ref{eq:partition_f1}), we then get
 \be \label{eq:lfd1}
\mathcal{P}(s, N) \sim e^{-N^3\Psi(s)} \quad \; {\rm where} \quad \; \Psi(s) = S[\rho^*_{\mu}(x), \mu, \mu_0] + \frac{2}{3}\alpha^2 \;.
\ee
By substituting the explicit solution for the saddle-point density from Eq. (\ref{eq:second_xB}), we find after straightforward algebra,  
\begin{align}\label{eq:LDF}
    \Psi(s)&=\frac{L_2^3-L_1^3}{48\alpha}+\frac{L_2^2-L_1^2}{8}+\frac{L_2^2}{4}-\alpha L_2+\frac{2}{3}\alpha^2+\nonumber \\
 + \frac{\mu}{4}&\left[\frac{1}{4\alpha}\int_{L_1}^{L_2}x^2f''(x)dx -2s +2f(L_2)+\int_{L_1}^{L_2}xf''(x)dx \right] \;,
\end{align}
where $L_1, L_2$ and $\mu$ are functions of $s$ and are determined from the relations (\ref{eq:L2}), (\ref{eq:L1}), (\ref{eq:mu0_2}) and (\ref{eq:third}). This expression gives the exact rate function for the linear statistics $s = (1/N)\sum_i f(x_i)$ with arbitrary $f(x)$. This is one of the main results of this paper.  
Let us end this subsection with a remark. When $s$ approaches $\bar{s}$ given in Eq. (\ref{av_s}), we have seen that $\mu \to 0$ [see the discussion below Eq. (\ref{eq:third})] and the density $\rho_\mu^* (x)\to \rho_0^*(x) = 1/(4\alpha)$ for $L_1\leq x \leq L_2$ with $L_1 \to -2 \alpha$ and $L_2 \to 2 \alpha$. Consequently, one can easily verify from Eq. (\ref{eq:lfd1}) that $\Psi(s) \to 0$ as $s \to \bar{s}$. Since $\Psi(s)$ is a positive convex function, clearly $s = \bar{s}$ is a minimum of $\Psi(s)$.

\subsection{Extraction of the variance in the large $N$ limit}

In the previous section, we have determined the large deviation form of the PDF ${\cal P}(s,N)$ in Eq. (\ref{eq:lfd1}) with the rate function $\Psi(s)$ given in Eq. (\ref{eq:LDF}). The goal of this subsection is to extract the variance by expanding the rate function $\Psi(s)$ around its minimum at $s = \bar{s}$, as explained in the beginning of this section. We have shown at the end of the previous subsection that, indeed, $\Psi(s)$ has a minimum at $s = \bar{s}$ with $\bar{s}$ given in Eq. (\ref{av_s}). For easy reading, we recall it here
\bea \label{av_s2}
\bar{s} = \frac{1}{4\alpha}\int_{-2\alpha}^{2\alpha} f(x)\,dx \;.
\eea
We now expand $\Psi(s)$ around this minimum at $s=\bar{s}$. We have seen that when $s = \bar{s}$, the saddle-point solution is the unconstrained flat density over $[L_1, L_2]$ with $L_1 = -2 \alpha$ and $L_2 = +2 \alpha$. When $s$ changes slightly from $\bar{s}$, say $s = \bar{s} + \epsilon$, we expect 
\begin{align}\label{eq:around_avr}
    s = \bar{s} + \epsilon \;, \; L_1 = -2\alpha +\delta_1(\epsilon) \;, \; L_2 = 2\alpha + \delta_2(\epsilon) \;,
\end{align}
where $\delta_1(\epsilon)$ and $\delta_2(\epsilon)$ are small. The idea is first to evaluate $\mu, L_1$ and $L_2$ using the independent relations (\ref{eq:L2}), (\ref{eq:L1}) and (\ref{eq:third}) for small $\epsilon$ and then substitute this result in Eq. (\ref{eq:LDF}) and evaluate $\Psi(s)$ up to order $O(\epsilon^2)$, where we recall that $\epsilon = s-\bar{s}$. We also recall that $\Psi(s)$ is just the saddle point action up to a constant -- see Eq. (\ref{eq:lfd1}). It turns out that, in order to extract the expansion of $\Psi(s)$ around $\bar{s}$, it is enough to compute $\mu$ only up to order $O(\epsilon)$. This follows from a thermodynamic identity \cite{Grabsch1,Grabsch4,Cunden_s} (for a simple proof, see Eqs. (40)-(41) in Ref. \cite{Flack2021}) which states that
\bea \label{thermo_id}
\frac{\partial \Psi}{\partial s} = \frac{\partial S[\rho^*_{\mu}(x), \mu, \mu_0] }{\partial s} = - \mu(s) \;,
\eea
where the first equality follows from Eq. (\ref{eq:lfd1}). From this relation (\ref{thermo_id}) it is clear that to expand $\Psi(s)$ around $\bar{s}$ up to quadratic order, it is enough to compute $\mu(s)$ only up to linear order in $\epsilon = s-\bar{s}$. 
We then substitute the expansions (\ref{eq:around_avr}) in the four independent relations (\ref{eq:L2}), (\ref{eq:L1}), (\ref{eq:mu0_2}) and (\ref{eq:third}) and expand up to order $\epsilon$. This straightforward expansion leads to
\bea  \label{mu_small_eps}
\mu = -  \frac{4\alpha \, \epsilon}{\int_{-2\alpha}^{+2 \alpha} [f'(x)]^2\,dx} + {O}(\epsilon^2)\;.
\eea 
In addition, $\delta_1(\epsilon)$ and $\delta_2(\epsilon)$ are given by
\bea 
&&\delta_1(\epsilon)  = \frac{4 \alpha f'(-2 \alpha) \, \epsilon}{\int_{-2\alpha}^{+2 \alpha} [f'(x)]^2\,dx} + {O}(\epsilon^2) \label{delta1_small_eps} \\
&&\delta_2(\epsilon)  = \frac{4 \alpha f'(2 \alpha) \, \epsilon}{\int_{-2\alpha}^{+2 \alpha} [f'(x)]^2\,dx} + {O}(\epsilon^2) \;. \label{delta2_small_eps}
\eea
We then substitute this relation (\ref{mu_small_eps}) in (\ref{thermo_id}) and integrate over $s$ to obtain
\bea \label{exp_psi}
\psi(s) = \frac{1}{2b}(s-\bar{s})^2 + O((s-\bar{s})^3) \quad, \quad {\rm with} \quad b = \frac{1}{4\alpha}\int_{-2\alpha}^{2 \alpha} [f'(x)]^2 \, dx \;.
\eea
Hence, from Eq. (\ref{rel_b_var}), we get
\bea \label{var_explicit}
{\rm Var}(s) \approx \frac{b}{N^3} \approx  \frac{1}{4 \alpha N^3} \int_{-2 \alpha}^{2 \alpha}  \left[f'(x)\right]^2 \, dx  \;.
\eea

\subsection{A few examples}

Here we work out two simple examples of $f(x)$, namely $f(x) = x$ and $f(x)  = x^2$.

%
%
%
%
%

\begin{figure}[t]
\centering
\includegraphics[width = \linewidth]{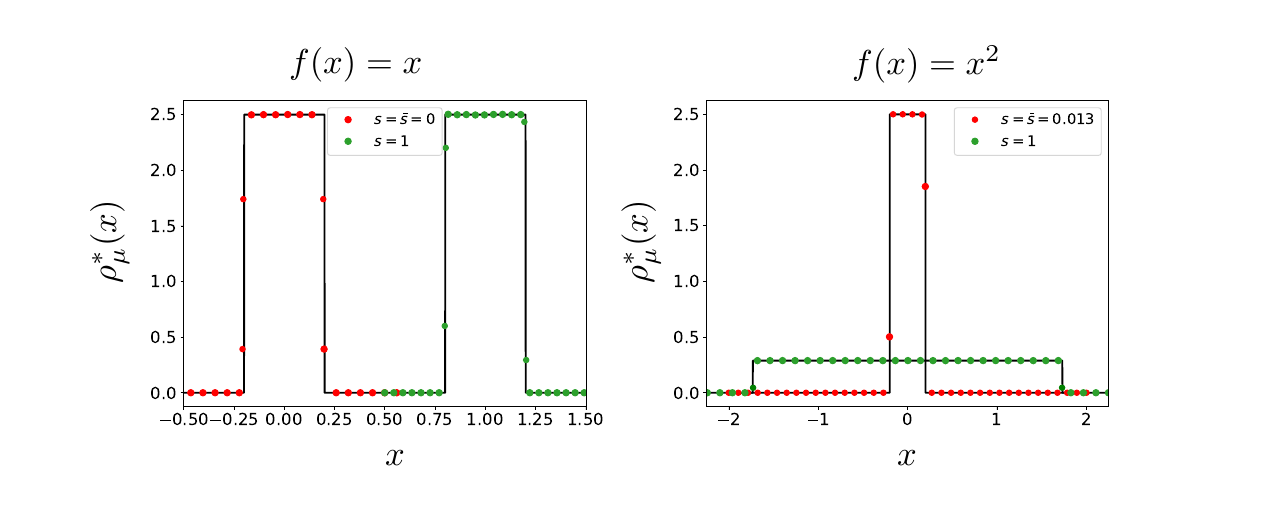}
\caption{{\bf Left:} a plot of the saddle-point density for $f(x)=x$ and two different values of $s=\bar{s}=0$ and $s=1$. For $f(x) = x$, the average unperturbed value of $s$ is $\bar{s}=0$, where the density is flat over $[-2\alpha, 2 \alpha]$ (this is shown by the red curve). When $s$ changes from $s=0$ to $s=1$, the new density (shown by the green curve) gets shifted by $s=1$. {\bf Right:} a plot of the saddle-point density for $f(x)=x^2$ and two different values of $s=\bar{s}=4 \alpha^2/3$ and $s=1$. For $f(x) = x^2$, the average unperturbed value of $s$ is $\bar{s}=4 \alpha^2/3$, where the density is flat over $[-2\alpha, 2 \alpha]$ (this is shown by the red curve). When $s$ changes from $s=\bar{s}$ to $s=1$, the new density (shown by the green curve), while remaining flat over $[-\sqrt{3},+\sqrt{3}]$, gets rescaled (unlike in the case $f(x) = x$ where it is just shifted without changing the shape from the unperturbed case). In the figures, we chose $\alpha = 1/10$.} \label{Fig_x}
\end{figure}
\begin{figure}[t]
\centering
\includegraphics[width = 0.6\linewidth]{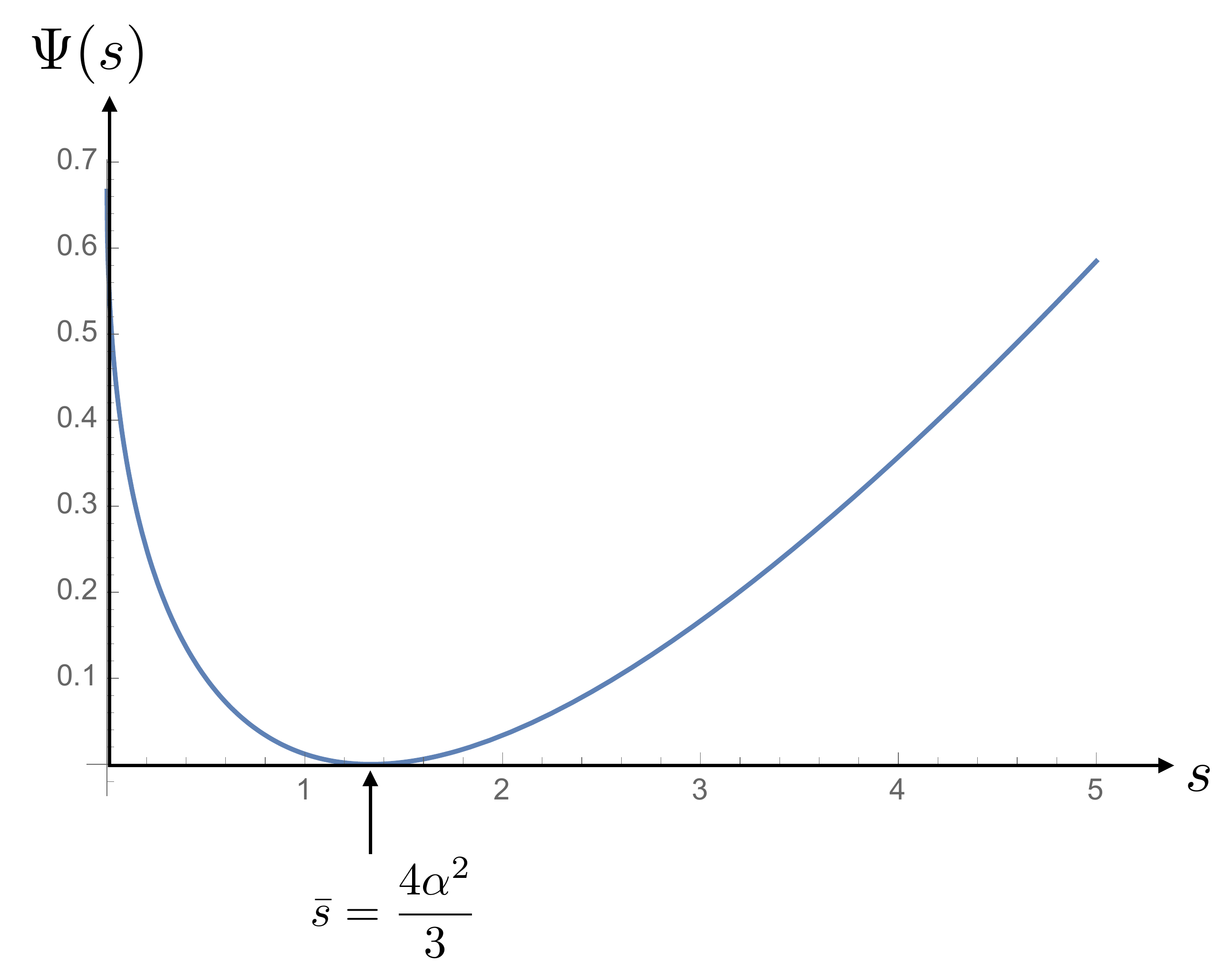}
\caption{Plot of the large deviation $\Psi(s)$ vs $s$, for $f(x) = x^2$ as given in Eq. (\ref{psi_neq2}), with the choice $\alpha = 1$.}\label{Plot_Psi}
\end{figure}

\begin{itemize}
\item{}{$f(x) = x$: in this case the effective potential in Eq. (\ref{Veff}) reads $V_{\rm eff}(x) = x^2/2 + \mu x$, which is always confining for all $\mu$ and hence we expect a single support solution for the density, an assumption which is crucial for the derivation of the general formula for the variance of $s$ in Eq. (\ref{var_explicit}). In this case, the saddle-point density in Eq. (\ref{eq:second_xB}) reads
\bea \label{rhoneq1}
\rho_\mu^*(x) = \frac{1}{4\alpha} {\mathbb I}_{[L_1, L_2]}(x)  \;,
\eea
where ${\mathbb{I}}_{[L_1,L_2]}(x)$ is an indicator function, which is $1$ if $x \in [L_1, L_2]$ and $0$ otherwise. 
The parameters $L_1$, $L_2$ and $\mu$ are determined respectively from Eqs. (\ref{eq:L1}), (\ref{eq:L2}) and (\ref{eq:third}). They simply read
\bea \label{param_neq1}
L_1 = s-2 \alpha \quad, \quad L_2 = s+2\alpha \quad, \quad \mu = -s \quad {\rm and} \quad \bar{s} = 0 \;.
\eea
Thus the original density $\rho_0^*(x) = \frac{1}{4\alpha}\,\mathbb{I}_{[-2\alpha,2\alpha]}(x)$ just gets shifted for a finite $s$, or equivalently for finite $\mu$, as seen from Eq. (\ref{rhoneq1}) (see the left panel of Fig. \ref{Fig_x}). Since $\mu(s) = -s$ for all $s$, we can integrate the exact relation (\ref{thermo_id}) with the condition that $\Psi(s=\bar{s}) = 0$. This gives 
the full rate function  
\bea \label{psi_neq1}
\Psi(s) = \frac{s^2}{2} \;. 
\eea
Hence, clearly, from Eq. (\ref{def_psi}), it follows that the variance is ${\rm Var}(s) \approx b/N^3$ with $b=1$, in agreement with our general formula in Eq. (\ref{var_explicit}).
}
\item{}{$f(x) = x^2$: in this case the effective potential in Eq. (\ref{Veff}) reads
$V_{\rm eff}(x)  = (\mu + 1/2)x^2$. Hence, for all $\mu > -1/2$, the potential is confining and we expect to have a single support around $x=0$. The 
saddle-point density from Eq. (\ref{eq:second_xB}) reads
\bea \label{rhoneq2}
\rho_\mu^*(x) = \frac{1}{2 \sqrt{3\,s}} {\mathbb I}_{[L_1, L_2]}(x) \;,
\eea
with the parameters $L_1$, $L_2$ and $\mu$ again determined respectively from Eqs. (\ref{eq:L1}), (\ref{eq:L2}) and (\ref{eq:third}). We get
\begin{equation} \label{param_neq2}
L_1 = - \sqrt{3\,s} \quad, \quad L_2 = \sqrt{3\,s} \quad, \quad \mu = -\frac{1}{2} + \frac{\alpha}{\sqrt{3\,s}} \quad {\rm and} \quad \bar{s} = \frac{4 \alpha^2}{3} \;.
\end{equation}
Note that, for any $s>0$, we have $\mu > -1/2$ from Eq. (\ref{param_neq2}) and hence the effective potential $V_{\rm eff}(x) = (\mu + 1/2)x^2$ will be always confining for all $s$, leading to a single support solution for any $s$. 
Thus, as in the $f(x) = x$ case discussed before, the saddle-point density is uniform but unlike the case $f(x) = x$, it is not just a shift of $\rho_0^*(x)$ but also the height
and the width gets modified (see the right panel of Fig. \ref{Fig_x}). Using the exact $\mu(s)$ from Eq. (\ref{param_neq2}) in the exact relation (\ref{thermo_id}), and integrating with respect to $s$ using $\Psi(s= \bar{s}) = 0$, one gets the full rate function 
\bea \label{psi_neq2}
\Psi(s) = \frac{s}{2} - \frac{2\alpha}{\sqrt{3}} \sqrt{s} + \frac{2\alpha^2}{3} \;.
\eea
A plot of this rate function is shown in Fig. \ref{Plot_Psi}. By expanding around $s = \bar{s} = 4\alpha^2/3$ up to quadratic order, one gets ${\rm Var}(s) =b/N^3$ with $b= 16\,\alpha^2/3$, in agreement with our general formula in Eq. (\ref{var_explicit}).

}
\end{itemize}

Note that in the derivation of the general formula (\ref{var_explicit}), we have assumed that the effective potential $V_{\rm eff}(x) = x^2/2 + \mu f(x)$ is confining, so that one has a single support for the saddle-point density and also the fact that $f(x)$ is a smooth function such that its first derivative $f'(x)$ exists and the integral in Eq. (\ref{var_explicit}) is finite. In the two examples discussed above, namely $f(x) = x$ and $f(x) = x^2$, both conditions are met. In Section~\ref{sec:validity}, we will demonstrate examples where either one of the two conditions breaks down and yet we will show that the formula (\ref{var_explicit}) will still be valid.

\section{An alternative derivation of the formula for the variance} \label{sec:alter}

%
%
%

In this section, we provide an alternative derivation of the formula for the variance given in Eq. (\ref{var_s}). Following Ref. \cite{Joh} for the log-gas case, we focus on the generating function of $s$, namely
\bea \label{def_G}
G(\lambda) = \langle e^{- \lambda s }\rangle = \langle e^{- \frac{\lambda}{N} \sum_{i=1}^N f(\lambda_i)}\rangle  = \int_{-\infty}^{\infty} ds \, {\cal P}(s,N) e^{- \lambda s} \;,
\eea 
where we recall that $\langle \cdots \rangle$ denotes an average over the Gibbs-Boltzmann measure in Eq.~(\ref{gibbs}) and $ {\cal P}(s,N)$ is the PDF of $s$ given in Eq. (\ref{eq:LS0}). The starting point of the analysis presented here is the following identity
\bea \label{id_G}
\frac{\partial}{\partial \lambda} \ln G(\lambda) = - \frac{1}{N} \Big \langle \sum_{i=1}^N f(x_i) \Big \rangle_\lambda \;,
\eea
where $\langle \cdots \rangle_\lambda$ denotes an average with respect to the modified weight
\bea \label{mod_gibbs}
\tilde P_\lambda (x_1, x_2, \cdots, x_N) = \frac{1}{\tilde Z_N(\lambda)} \, e^{- E[\{ x_i\}] - \frac{\lambda}{N} \sum_{i=1}^N f(x_i)} \;,
\eea
where $\tilde Z_N(\lambda)$ is a normalisation constant. Hence, in the large $N$ limit the relation (\ref{def_G}) becomes, to leading order for large $N$
\bea \label{id_G2}
\frac{\partial}{\partial \lambda} \ln G(\lambda) \approx - \int_{-\infty}^\infty \tilde \rho_\lambda(x) f(x) \, dx \;,
\eea
where $\tilde \rho_\lambda(x)$ is the equilibrium density associated to the joint PDF in (\ref{mod_gibbs}). By comparing Eqs. (\ref{mod_gibbs}) and (\ref{eq:ener2}), it is easy to see that, to leading order for large $N$,   
\bea \label{rel_lambda_mu}
\tilde \rho_\lambda(x) \approx \rho_{\mu = \lambda/N^3}^*(x) \;,
\eea
where $\rho_\mu^*(x)$ is given in Eq. (\ref{eq:second_xB}). Hence $\tilde \rho_\lambda(x)$ reads, to leading order for large $N$,
\bea \label{rho_tilde}
\tilde \rho_\lambda(x) \approx \frac{1}{4 \alpha} \left( 1 + \frac{\lambda}{N^3}f''(x)\right) \;, \; \tilde L_1(\lambda) \leq x \leq \tilde L_2(\lambda) \;,
\eea
where $\tilde L_1(\lambda)$ and $\tilde L_2(\lambda)$ are given respectively by Eqs. (\ref{eq:L1}) and (\ref{eq:L2}) with the substitution $\mu  = \lambda/N^3$, i.e.,
\bea \label{eq_ltilde}
&&\tilde L_1(\lambda) + 2 \alpha + \frac{\lambda}{N^3} f'(\tilde L_1(\lambda)) = 0 \label{eq_ltilde1} \\
&&\tilde L_2(\lambda) - 2 \alpha + \frac{\lambda}{N^3} f'(\tilde L_2(\lambda)) = 0 \label{eq_ltilde2}
\eea
To compute the variance, we need to compute the small $\lambda$ expansion of $\ln G(\lambda)$ in (\ref{id_G2}) up to order $O(\lambda^2)$. For this purpose, it turns out that we need to expand $\tilde L_1(\lambda)$ and $\tilde L_2(\lambda)$ up to order $O(\lambda)$ only. Expanding Eqs. (\ref{eq_ltilde1}) and (\ref{eq_ltilde2}) for small $\lambda$ one finds
\be  \label{tildeL_largeN}
\tilde L_1(\lambda) = -2 \alpha - \frac{\lambda}{N^3} f'(2 \alpha) + O(\lambda^2) \quad, \quad \tilde L_2(\lambda) = 2 \alpha - \frac{\lambda}{N^3} f'(2 \alpha) + O(\lambda^2) \;.
\ee
Inserting Eq. (\ref{rho_tilde}) in Eq. (\ref{id_G2}) one obtains 
\bea \label{id_G3}
\frac{\partial}{\partial \lambda} \ln G(\lambda) \approx - \frac{1}{4 \alpha}\int_{-\tilde L_1}^{\tilde L_2}  f(x) \, dx - \frac{\lambda}{4 \alpha \, N^3}\int_{\tilde L_1}^{\tilde L_2} dx \, f(x) f''(x) \;.
\eea
Performing an integration by parts in the second integral in (\ref{id_G3}) and using the small $\lambda$ expansion of $\tilde L_1(\lambda)$ and $\tilde L_2(\lambda)$ in (\ref{tildeL_largeN}), one obtains after straightforward algebra
\bea \label{id_G4}
\frac{\partial}{\partial \lambda} \ln G(\lambda) = - \frac{1}{4\alpha} \int_{-2 \alpha}^{2 \alpha} f(x) \, dx + \frac{\lambda}{4 \alpha N^3} \int_{-2 \alpha}^{2 \alpha} dx [f'(x)]^2  + O(\lambda^2)\;.
\eea
Therefore, integrating over $\lambda$, using $G(\lambda = 0)=1$ one finds
\bea \label{id_G5}
\ln G(\lambda) = - \frac{\lambda}{4 \alpha} \int_{-2 \alpha}^{2 \alpha} f(x) \, dx + \frac{\lambda^2}{8 \alpha N^3} \int_{-2 \alpha}^{2 \alpha} dx [f'(x)]^ 2 + O(\lambda^3) \;.
\eea
From this expression, one can immediately read off the mean $\bar{s}$ and the variance ${\rm Var}(s)$ of $s$ as
\bea \label{cumul}
\bar{s} = \frac{1}{4 \alpha}  \int_{-2 \alpha}^{2 \alpha} f(x) \, dx \quad, \quad {\rm Var}(s) =  \frac{1}{4 \alpha N^3} \int_{-2 \alpha}^{2 \alpha} dx \, [f'(x)]^ 2 \;,
\eea
which indeed coincide with the results in Eqs. (\ref{av_s}) and (\ref{var_s}) obtained by a different method. 

The exercice above can actually be repeated for arbitrary confining potential $V(x)$, not necessarily harmonic. In this case, the equilibrium density in Eq. (\ref{eq:second_xB}) gets replaced by
\bea \label{rho_V}
\tilde \rho_\mu^*(x) = \frac{1}{4\alpha} \left[ V''(x) + \mu f''(x) \right] \;,
\eea
which is supported over the interval $[L_1,L_2]$. Note that $L_1$ and $L_2$ do depend on $V(x)$. 
One can then repeat the steps above and finds that the formula for the variance (\ref{cumul}) actually holds for a general $V(x)$ and reads, up to an overall $N$-dependent constant,
\bea \label{var_genV}
{\rm Var}(s) \propto \int_{L_1}^{L_2} dx \, [f'(x)]^ 2 \;.
\eea  
Thus the dependence of the variance on $V(x)$ enters only through the support edges $L_1$ and $L_2$, but not explicitly.

\section{Validity of the formula for the variance} \label{sec:validity}

As mentioned above, one of the crucial assumptions leading to the derivation of the formula (\ref{var_explicit}) is the following. Once the chemical potential is switched on, it changes the effective potential of the jellium to $V_{\rm eff}(x) = x^2/2 + \mu f(x)$ [see Eq. (\ref{Veff})]. 
We assume $V_{\rm eff}(x)$ is still confining and the saddle-point density still has a single support. In addition $f'(x)$ must exist and the integral in Eq. (\ref{var_explicit}) should be finite. Below, we discuss two examples where one of the two conditions breaks down. For instance, when $f(x) = x^3$, the effective potential is always non confining, except of course for $\mu=0$. In some other examples, $f'(x)$ maybe be singular, such as $f(x) = |x|$, which also leads to two disjoint supports for the density
for $\mu <0$. However, in both cases, we will show that the general formula (\ref{var_explicit}) still gives the correct answer for the variance of $s$.

%
%

\subsection{The special case $f(x) = x^3$}

In this case, the effective potential $V_{\rm eff}(x) = x^2/2 + \mu x^3$ is non-confining for any $\mu$. For example, for $\mu >0$, 
it diverges negatively as $x \to -\infty$. A plot of $V_{\rm eff}(x$ in this case is given in Fig. \ref{fig_xcube} for $\mu>0$. 
\begin{figure}[t]
\centering
\includegraphics[width = 0.7\linewidth]{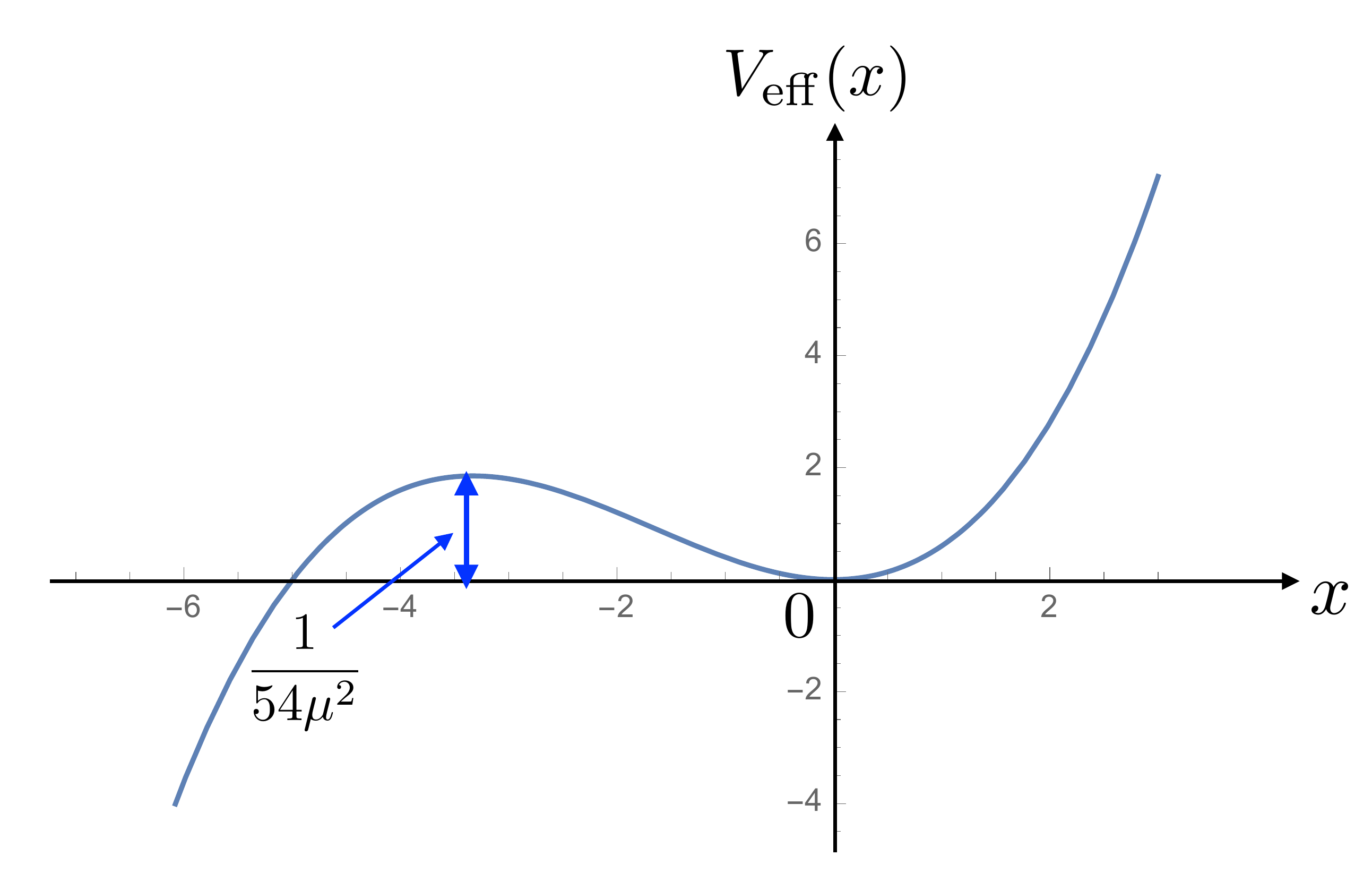}
\caption{Plot of the effective potential $V_{\rm eff}(x) = x^2/2 + \mu x^3$ with $\mu = 1/10$. The maximum on the left of the origin occurs at $x = - 1/(3 \mu)$, while the height of the maximum is $1/(54 \mu^2)$.}\label{fig_xcube}
\end{figure}

For $\mu>0$, the effective potential 
$V_{\rm eff}(x)$ has always a minimum at $x=0$ and a maximum at $x = -1/(3 \mu)$. The height of the maximum is given by
\bea \label{height}
V_{\rm eff}\left( x = - \frac{1}{3\mu}\right) = \frac{1}{54\, \mu^2} \;.
\eea
For small $\mu$, the barrier height is thus very large and, hence, the minimum at $x=0$ is deep. Hence, for small $\mu$, we expect to have
a single support. For large positive $\mu$, the barrier height will decrease and the minimum at $x=0$ will be shallow. In this case, some charges may split from the support around $x=0$ and go over the barrier to very large negative $x$ and our assumption about the single support will no
longer be valid. However, for the computation of the variance of $s$, we only need $\mu$ very small. Hence, the single support assumption will still be correct and hence we expect that our formula in Eq. (\ref{var_explicit}) is still valid for large $N$. Indeed, assuming a single support solution as in Eq. (\ref{eq:second_xB}), one can in principle compute the parameters $L_1, L_2$ and $\mu$ and check that the rate function $\Psi(s)$ in Eq. (\ref{eq:LDF}) is again quadratic around $s = \bar{s}$ with the variance given by Eq. (\ref{var_explicit}) with $b$ given by
\bea \label{b_cubic}
b = \frac{1}{4\alpha} \int_{-2\alpha}^{2 \alpha} (3 x^2)^2 \, dx = \frac{144}{5}\,\alpha^4 \;.
\eea
Thus, while we are able to compute the variance, from the expansion of $\Psi(s)$ in Eq. (\ref{eq:LDF}) around $s = \bar{s}$, the expression of the rate function $\Psi(s)$ --obtained by assuming a single support -- is not expected to be valid for all $s$. Computing the full rate function $\Psi(s)$ for all $s$ is an interesting challenge left for future investigations. Note that, here, we just discussed the specific example of $f(x) = x^3$, but a similar
discussion will hold for any odd function $f(x)$ that diverges faster than $x^2$, for instance $f(x) = x^5, x^7, \cdots$. In all such cases, we expect the formula for the variance in Eq. (\ref{var_explicit}) to be still valid, since it arises only from the small $\mu$ expansion. We have verified the validity of our prediction for the variance in (\ref{b_cubic}) for $f(x) = x^3$ by Monte-Carlo simulations, as shown in Fig. \ref{fig_xcube} where we plot the difference between the theoretical value of $b = \frac{144}{5}\,\alpha^4$ [see Eq. (\ref{b_cubic})] and the Monte-Carlo value $b_{\rm MC} = N^3 {\rm Var}(s)\Big|_{\rm MC}$ as a function of increasing $N$. We find that it decreases to zero as $N \to \infty$, thus verifying the theoretical prediction.  
\begin{figure}[t]
\centering
\includegraphics[width =0.7 \linewidth]{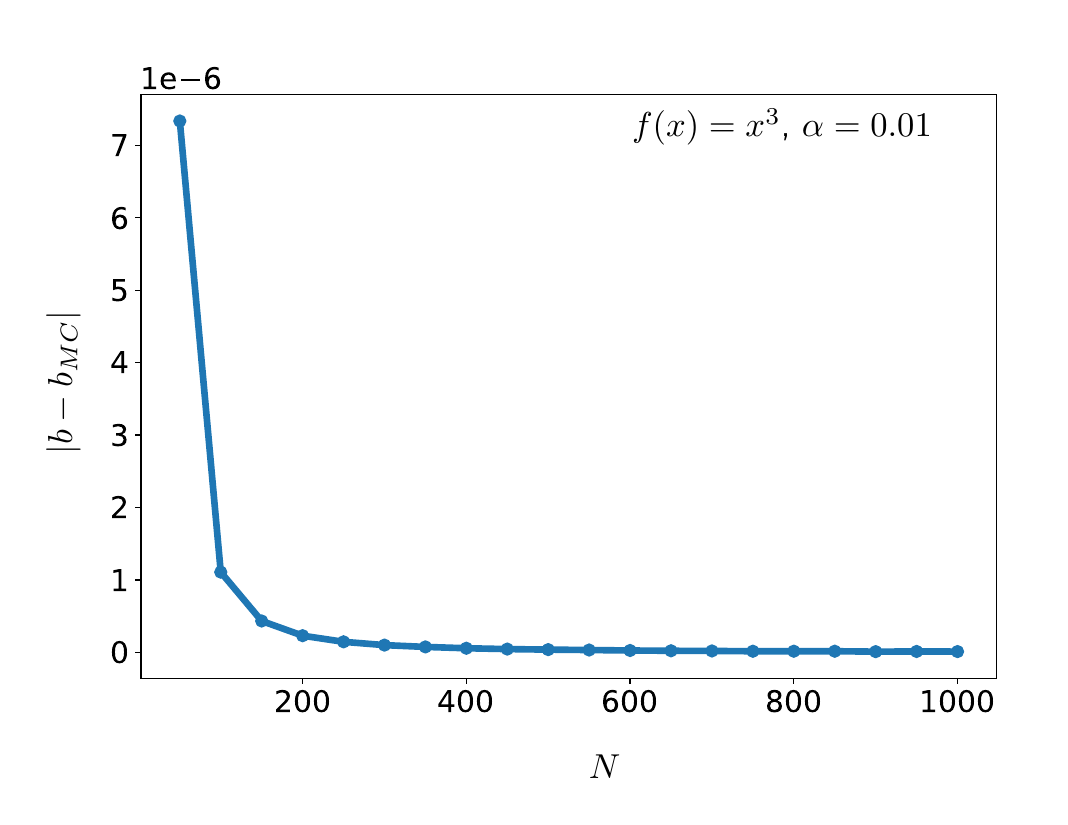}
\caption{A plot of $|b-b_{\rm MC}|$ vs $N$ where $b= \frac{144}{5}\,\alpha^4$ (with $\alpha = 0.01$) is the theoretical prediction for $N^3 {\rm Var}(s)$ for $f(x)=x^3$ and $b_{\rm MC}$ is the Monte-Carlo value of $N^3 {\rm Var}(s)$.}\label{fig_xcube}
\end{figure}

\subsection{The special case $f(x) = |x|$}

\begin{figure}[t]
    \centering
    \includegraphics[width=1\linewidth]{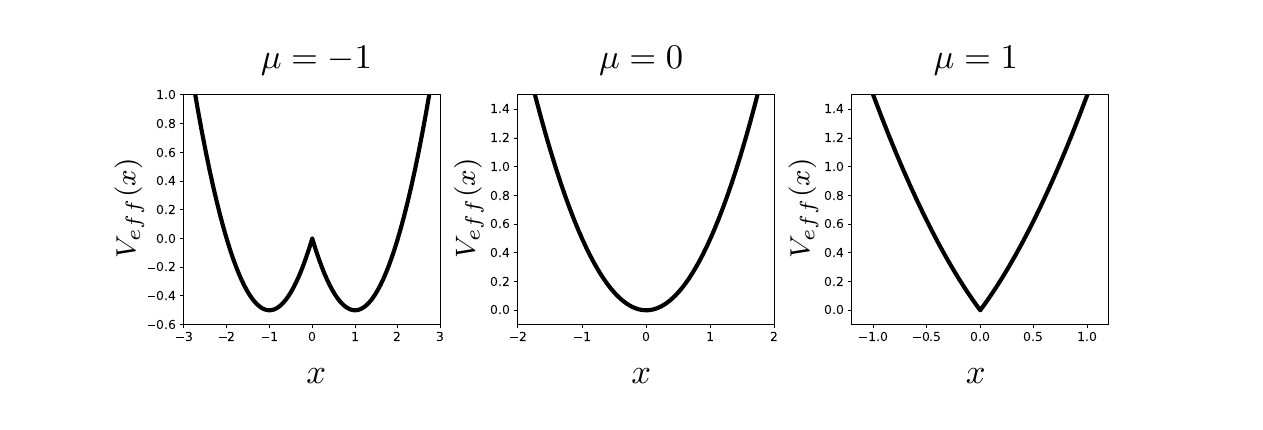}
    \caption{Effective potentials $V_{\rm eff}(x) = x^2/2 + \mu |x|$ vs $x$ for three values of $\mu$. From left to right the values of $\mu$ are $-1$, $0$, and $1$. If $\mu<0$ the potential has two minima and for $\mu>0$ just one. The transition point is at $\mu=0$.}
    \label{fig:potentials}
\end{figure}

In this case, the effective potential in Eq. (\ref{Veff}) felt by the charges in the jellium model reads
\bea \label{Veff_mod}
V_{\text{eff}}(x)= \frac{1}{2}x^2 + \mu \, |x| \;.
\eea
Clearly, for $\mu >0$, $V_{\rm eff}(x)$ has a single minimum at $x = 0$. In this case, one would expect that the saddle-point density $\rho^*_\mu(x)$ will be supported over a single interval around this minimum. In contrast, when $\mu <0$, the effective potential has two minima located at $\pm |\mu|$ (see Fig. \ref{fig:potentials}). 
Hence, in this case, one expects that the saddle-point density will have two disjoint supports, one around
each minimum. Below, we calculate the saddle-point density explicitly, confirming this scenario. Before proceeding, we note that the average value of $s$ (corresponding to $\mu=0$) is given by [using $f(x) = |x|$ in Eq. (\ref{av_s})]
\bea \label{sbar_mod}
\bar{s} = \alpha \;.
\eea

Our starting point is the action $S[\rho(x), \mu, \mu_0]$ in Eq. (\ref{eq:act2B}) with $f(x) = |x|$. The saddle point equation for the density then reads, from Eq. (\ref{eq:first_x})
\be \label{eq:first_x_mod}
x-2\alpha \int\rho^*_{\mu}(y){\rm sgn}(x-y)dy + \mu \,{\rm sgn}(x) = 0 \;.
\ee
Note that this equation holds only at points inside the support of $\rho^*_\mu(x)$. We now consider the two cases $\mu>0$ and $\mu<0$ separately.

\begin{figure}[t]
    \centering
    \includegraphics[width=1\linewidth]{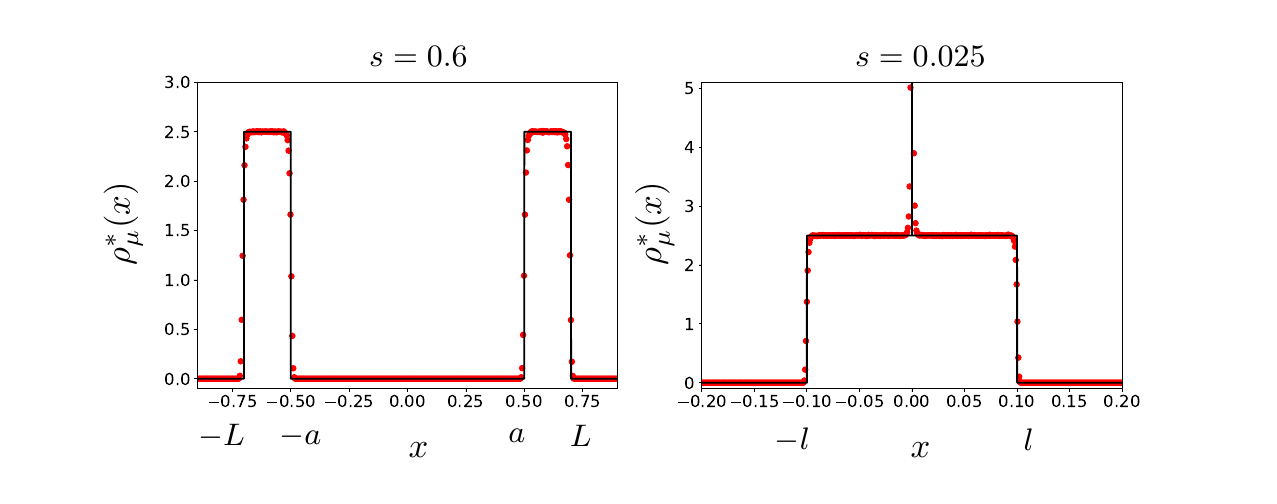}
    \caption{{\bf Left:} Plot of the Monte-Carlo simulations for the average density (red dots) for $\alpha =1/10$ and $N=1000$ for $s = 0.6$, compared with the saddle-point prediction in Eq. (\ref{rho_bulk}). In this case, since $s = 0.6 > \bar{s} = \alpha = 1/10$, the density has two disjoint supports with parameters $L = s + \alpha = 0.7$ and $a = s - \alpha=0.5$. {\bf Right:} Plot of the Monte-Carlo simulations for the average density (red dots) for $\alpha =1/10$ and $N=1000$ for $s = 0.025$, compared with the saddle-point prediction in Eq. (\ref{rho_sp_mod}). In this case since $s =0.025 < \bar{s} = 1/10$, the density has a single support with an additional delta peak at $x=0$.}\label{fig:ansatzs}
\end{figure}

\vspace{0.3cm}
\noindent{\bf The case $\mu >0$:} In this case, we expect that there is a single support over $[-\ell, + \ell]$ where $\ell$ remains to be determined. 
Note that, since $V_{\rm eff}(x)$ is symmetric around $x=0$, we expect the support to be also symmetric around $x=0$, and hence we chose it to be $[-\ell, + \ell]$. Taking one more derivative of Eq. (\ref{eq:first_x_mod}) with respect to $x$, and using $\frac{d}{dx}\,{\rm sgn}(x) = 2 \delta(x)$, we get
\bea \label{rho_sp_mod}
\rho_\mu^*(x) = \frac{1}{4\alpha} + \frac{\mu}{2\alpha}\, \delta(x) \;.
\eea   
Thus the density has a flat profile with a spike (delta-function) at its center. This is confirmed in our Monte-Carlo simulations (see the middle panel Fig \ref{fig:ansatzs}). The unknown parameters are $\ell$ and $\mu$. The normalisation condition $\int_{-\ell}^{+\ell}\rho_\mu^*(x)\, dx = 1$ gives the relation
\bea \label{ell_mu}
\ell +\mu = 2\alpha \;,
\eea 
and the other condition $s = \int_{-\ell}^{\ell} \rho(x) |x|\, dx$ gives
\bea \label{mu_s_mod}
s = \frac{1}{4\alpha} (2\alpha - \mu)^2 \;.
\eea
By inverting this relation, one gets $\mu = 2\alpha \pm \sqrt{4\,\alpha\,s}$. From Eq. (\ref{ell_mu}), it is clear that $\mu < 2 \alpha$ in order that
$\ell > 0$. Hence we choose the negative root of the quadratic equation for $\mu$ and set
\bea \label{mu_mod}
\mu = 2 \alpha - \sqrt{4 \,\alpha\,s} \quad {\rm and} \quad \ell = \sqrt{4 \alpha s} \;.
\eea
Since $\mu >0$, it follows that this single support solution in Eq. (\ref{eq:first_x_mod}) is valid only for $s < \alpha = \bar{s}$, where we used Eq. (\ref{sbar_mod}) for the last equality.

\vspace{0.3cm}
\noindent{\bf The case $\mu <0$:} In this case, as mentioned before, we expect a two-support solution. We need to parametrize the solution and we expect the two supports to be symmetrically placed symmetrically around the origin. Let us assume that the supports are $[-L,-a] \cup [a,L]$ with $0\leq a \leq L$ -- see Fig. \ref{fig:ansatzs}. Taking a derivative with respect to $x$ in the general saddle-point equation (\ref{eq:first_x_mod}), we see that the density in the bulk (i.e., away from the edges) is always flat and is given by 
\bea \label{rho_bulk}
\rho_\mu^*(x) = \frac{1}{4 \alpha} \quad {\rm for} \quad -L < x < -a \quad {\rm or} \quad a<x<L \;.  
\eea
Note that Eq. (\ref{eq:first_x_mod}) holds for all points belonging to both supports. Choosing $x \in [a,L]$ (right support), and using (\ref{rho_bulk}), one gets, from Eq. (\ref{eq:first_x_mod}) the relation 
\bea \label{mu_a1}
\mu = -a \;.
\eea
The normalisation condition $2\int_{a}^L \rho_\mu^*(x)\,dx = 1$ gives the relation
\bea \label{norm_mod}
L - a = 2 \alpha \;.
\eea
Furthermore, the condition $2 \int_a^L |x|\, \rho_\mu^*(x)\,dx = s$ gives another relation
\bea \label{s_mod}
L^2 - a^2 = 4 \alpha\, s \;.
\eea  
Thus we have three equations (\ref{mu_a1})-(\ref{s_mod}) for three unknowns $\mu, a$ and $L$ for a fixed $s$. 
Solving them, we get the three parameters 
\bea \label{3param}
L = s +\alpha \quad, \quad a = s - \alpha \quad {\rm and} \quad \mu = \alpha - s \;.
\eea
Since $\mu<0$, this two-support solution holds for $s > \alpha= \bar{s}$. 

Thus summarising the two cases $\mu >0$ (or equivalently $s<\bar{s}$) and $\mu < 0$ (or equivalently $s>\bar{s}$), we find that, as $s$ approaches $\bar{s}$ from above, the gap $2a = 2(s-\alpha) = 2(s - \bar{s})$ between the two supports shrinks linearly and vanishes exactly 
at $s = \bar{s}$, where the two supports merge with each other. When $s$ reduces further below $\bar{s}$, the delta-spike at $s=0$ appears
in the single support solution. Thus there is indeed a phase transition that occurs at $s = \bar{s}$ due to the vanishing of the gap between the two supports. This actually shows up in a singularity of the rate function $\Psi(s)$ at $s = \bar{s}$, as demonstrated below. 

In order to compute the rate function $\Psi(s)$ from the identity (\ref{thermo_id}), we need to first determine $\mu(s)$ as a function of $s$ for all $s$. Indeed, this is given from the analysis above in Eqs. (\ref{mu_mod}) and (\ref{3param}). We thus get 
\bea \label{mu_of_s}
\mu(s) = 
\begin{cases}
&2 \alpha - \sqrt{4 \alpha s} \quad, \quad s < \bar{s} = \alpha \\
& \\
& \alpha - s \quad \quad \quad \;, \quad s > \bar{s} = \alpha \;.
\end{cases}
\eea 
Substituting this in the identity (\ref{thermo_id}), integrating with respect to $s$, and using $\Psi(s = \bar{s}) = 0$, we get
\bea \label{Psi_of_s_mod}
\Psi(s) =
\begin{cases}
&- 2 \alpha\,s + \dfrac{4}{3} \sqrt{\alpha} s^{3/2} + \dfrac{2}{3}\alpha^2 \quad, \quad s < \alpha = \bar{s}  \;,\\
& \\
& \frac{1}{2}(s-\alpha)^2 \quad, \quad \hspace*{2.5cm} s > \alpha = \bar{s}  \;.
\end{cases}
\eea
One can easily check that $\Psi(s), \Psi'(s)$ as well as $\Psi''(s)$ are all continuous at $s = \bar{s} = \alpha$. However, the third derivative is discontinuous and is given by
\bea \label{Psi_third}
\Psi'''(s) =
\begin{cases}
&- \dfrac{1}{2\alpha} \quad, \quad s \to \bar{s}^-  \;,\\
& \\
& 0 \quad, \quad \hspace*{0.7cm} s \to \bar{s}^+  \;.
\end{cases}
\eea
\begin{figure}[t]
\centering
\includegraphics[width =0.7 \linewidth]{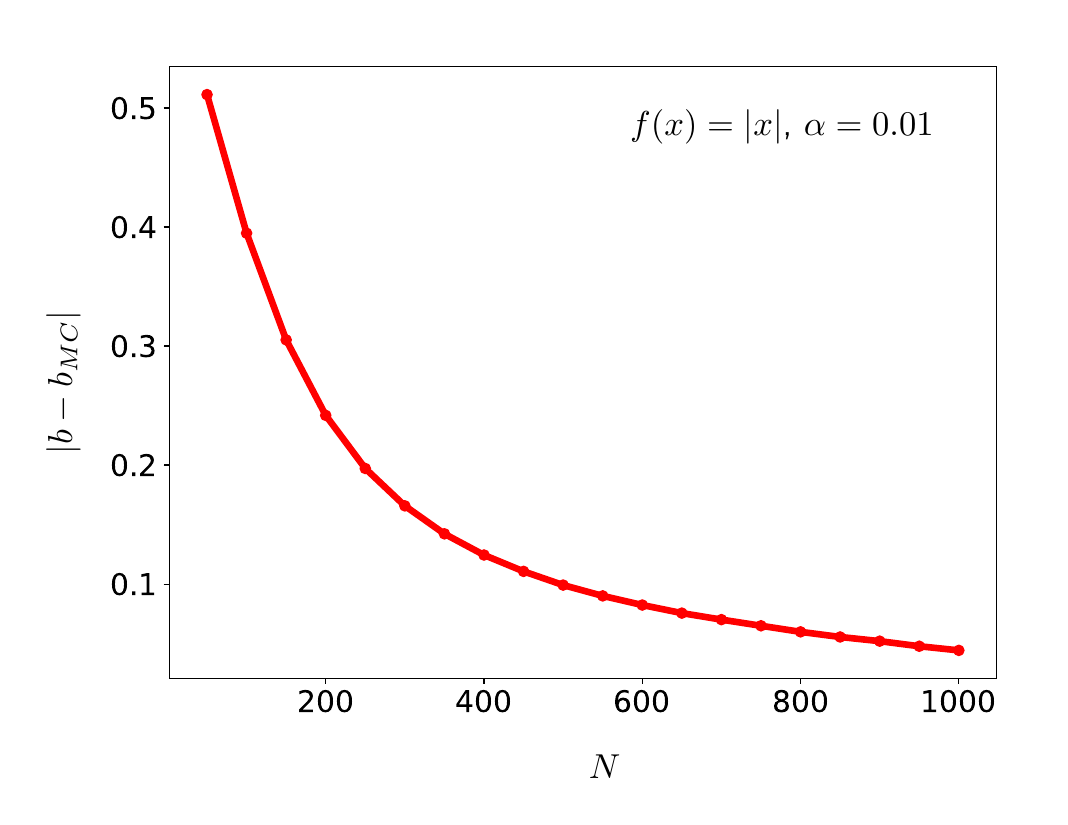}
\caption{A plot of $|b-b_{\rm MC}|$ vs $N$ where $b=1$ (with $\alpha = 0.01$) is the theoretical prediction for $N^3 {\rm Var}(s)$ for $f(x)=|x|$ and $b_{\rm MC}$ is the Monte-Carlo value of $N^3 {\rm Var}(s)$.}\label{fig_modx}
\end{figure}
Thus there is a third order phase transition in $\Psi(s)$ at $s = \bar{s}$. Such phase transitions have shown up in many other examples involving
Coulomb and log-gases (for a review see \cite{majumdar2014top}). Since $\Psi''(s)$ is continuous at $s = \bar{s}$, the expansion of $\Psi(s)$ around $s = \bar{s}$ is quadratic in leading order $\Psi(s) \approx \frac{1}{2}(s - \bar{s})^2$, indicating from Eq. (\ref{def_psi}) that
\bea \label{var_s_mod}
{\rm Var}(s) \approx \frac{b}{N^3} \quad {\rm with} \quad b = 1 \;.
\eea
Thus this is completely consistent with our general formula in Eq. (\ref{var_explicit}) which also predicts $b=1$ for $f(x) = |x|$. We have verified the validity of our prediction for the variance in (\ref{var_s_mod}) for $f(x) = x^3$ by Monte-Carlo simulations, as shown in Fig. \ref{fig_modx} where we plot the difference between the theoretical value of $b = 1$ (see Eq. (\ref{var_s_mod})) and the Monte-Carlo value $b_{\rm MC} = N^3 {\rm Var}(s)\Big|_{\rm MC}$ as a function of increasing $N$. We find that it decreases to zero as $N \to \infty$, thus verifying the theoretical prediction.

\section{Conclusion}\label{sec:final}

To summarize, we have considered the jellium model of $N$ particles in one-dimension with energy 
\bea \label{E_OCP_c}
 E[\{x_i\}] = \frac{N^2}{2}\sum_{i=1}^{N}x_i^2 - \alpha\, N\, \sum_{i\neq j}|x_i-x_j| \;,
\eea  
and studied the statistics of $s = (1/N) \sum_{i=1}^N f(x_i)$ in the Gibbs-Boltzmann state at any temperature of order $O(1)$. The main result in this paper is to derive a nice and compact general formula for the variance of $s$ in the limit of a large $N$
\be \label{eq:v_c}
{\rm Var}(s) = \frac{1}{N^3 4\alpha}\int_{-2\alpha}^{\alpha}(f'(x))^2 \, dx \;.
\ee
One expects this formula to be valid for a wide class of $f(x)$'s for which the integral in Eq. (\ref{eq:v_c}) is convergent. 
We have discussed with several examples, both analytically and numerically, the precise criteria behind the validity of this
formula. We have provided two different derivations of this formula in this paper. 

In addition to computing the formula for the variance for general $f(x)$, we have also shown that, for large $N$, the full PDF of $s$ exhibits a large deviation form ${\cal P}(s,N) \sim e^{-N^3 \Psi(s)}$ and we have computed the rate function $\Psi(s)$ for general $f(x)$ in Eq. (\ref{eq:LDF}) and provided more explicit forms in several examples such as $f(x) = x, x^2$ and $f(x) = |x|$. In all cases, $\Psi(s)$ exhibits a quadratic behavior around its minimum at $s = \bar{s} = (1/4\alpha) \int_{-2 \alpha}^{2 \alpha} f(x)\,dx$. This quadratic form reads $\Psi(s) \approx (s-\bar{s})^2/(2 {\rm Var}(s))$,  where ${\rm Var}(s)$ is given in Eq. (\ref{eq:v_c}).

It would be interesting to extend the technique presented in this paper to compute the covariance of two different linear statistics $s_1 = (1/N) \sum_{i=1}^N f(x_i)$ and $s_2 = (1/N) \sum_{i=1}^N g(x_i) $ where $f(x)$ and $g(x)$ are two different but arbitrary functions. In the log-gas case, this was computed in Ref. \cite{Cunden2014} and it is natural to compute this covariance of two linear statistics for the jellium model.

\ack
We thank P. Bourgade for useful discussions.


%
\end{document}